\documentclass[grl]{AGUTeX}
\usepackage{lineno}
\usepackage[dvips]{graphicx}
\setkeys{Gin}{draft=false}

\authorrunninghead{NOVOTRYASOV AND STEPANOV}

\titlerunninghead{INTENSIFICATION OF INTERNAL KELVIN WAVES}

\authoraddr{Vadim Novotryasov,
V.I.Il'ichev Pacific Oceanological Institute of the Russian Academy
of Sciences, 43 Baltiiskaya Street, 690041 Vladivostok, Russia.
(vadimnov@poi.dvo.ru)}

\authoraddr{Dmitriy Stepanov,
V.I.Il'ichev Pacific Oceanological Institute of the Russian Academy
of Sciences, 43 Baltiiskaya Street, 690041 Vladivostok, Russia.
(step-nov@poi.dvo.ru)}

\begin{document}

\title{Parametric interaction and intensification of nonlinear Kelvin waves}

\authors{Vadim Novotryasov and Dmitriy Stepanov}
\affil{V.I.Il'ichev Pacific Oceanological Institute, Vladivostok,
Russia}

\begin{abstract}
Observational evidence is presented for nonlinear interaction
between mesoscale internal Kelvin waves at the tidal -- $\omega _t$
or the inertial -- $\omega _i$ frequency and oscillations of
synoptic -- $\Omega $ frequency of the background coastal current of
Japan/East Sea. Enhanced coastal currents at the sum -- $\omega _+ $
and dif -- $\omega_-$ frequencies: $\omega_\pm =\omega_{t,i}\pm
\Omega$ have properties of propagating Kelvin waves suggesting
permanent energy exchange from the synoptic band to the mesoscale
$\omega _\pm $ band. The interaction may be responsible for the
greater than predicted intensification, steepen and break of
boundary trapped and equatorially trapped Kelvin waves, which can
affect El Ni\~{n}o. The problem on the parametric interaction of the
nonlinear Kelvin wave at the frequency $\omega $ and the
low-frequency narrow-band nose with representative frequency
$\Omega\ll\omega $ is investigated with the theory of nonlinear week
dispersion waves.
\end{abstract}

\begin{article}

\section{Introduction}
Internal Kelvin waves play a significant role in the dynamics of the
oceans. There are two basic types of the waves: equatorially trapped
and boundary trapped. Kelvin waves propagating on the equatorial
thermocline participate in the adjustment of the tropical ocean to
wind stress forcing [{\it Philander},1990]. Extensive data on Kelvin
waves have been obtained recently, motivated in part by possible
connection between the initial stages of El Ni\~{n}o and equatorial
nonlinear Kelvin waves which may precede this event. Any relaxation
or reversal of the steady trade winds (the easterlies) results in
the excitation of a Kelvin wave, which can affect El Ni\~{n}o. In
other words Kelvin waves play a critical role in generating and
sustaining of the Southern Oscillation [e.g.,{\it Fedorov}, 2000].

In the dynamics of the coastal oceans Kelvin waves are used to
interpret such phenomena as the instability of alongshore currents,
the generation and variability of wind currents on the shelf, and
upwelling [{\it Brink}, 1991]. {\it Fedorov and Melville} [1995,
1996] considered Kelvin waves trapped boundaries and showed that
waves could manifest nonlinear properties that is steepen and
overturn or break. A wave of depression deepens the thermocline and
breaks on the forward face of the wave. A broken wave of depression
can form a jump [also called shocks or fronts: {\it Lighthill},
1978; {\it Philander}, 1990]. In turn, the breaking of Kelvin waves
may be important in mixing, momentum and energy transfer in coastal
oceans. Other example of nonlinear Kelvin wave dynamics includes the
problem of nonlinear geostrophic adjustment in the presence of a
boundary [{\it Helfrich et al.}, 1999; {\it Reznik and Grimshaw},
2002].

The spectrum represents one of the major characteristics of the
waves. It is used as a representative statistical description of the
wave field in studies of nonlinear interaction [e.g., {\it Hibiya et
al.}, 1998], acoustic propagation [e.g., {\it Colosi et al.}, 1998],
and mixing parametrization [{\it Polzin}, 1995]. {\it Filonov and
Novotryasov} [2005, 2007] studded the wave band of temperature
fluctuation spectra in the coastal zone of Pacific Ocean and
observed that in the high-frequency band of temperature spectra the
spectral exponent tends to $\sim \omega^{-1}$ at the time of spring
tide and on the western shelf of the Japan/East Sea, in the $\omega
_i \ll\omega\ll N_\ast $ range, where $N_\ast $ is the
representative buoyancy frequency and $\omega _i $ is the inertial
frequency, the spectral exponent tends to $\sim \omega ^{-3}$. These
features {\it Filonov and Novotryasov} [2007] simulated by the model
spectrum of nonlinear internal waves in the shallow water. They
considered interaction of high-frequency waves with the wave at the
tidal frequency and shown that the spectrum of high-frequency
internal waves take the universal form and the spectral exponent
tends to $\sim \omega^{-1}$.

In this paper, we present observational evidence for nonlinear
interaction between synoptic oscillations of the background current
at the representative frequency $\Omega $ and nonlinear Kelvin waves
at the tidal -- $\omega _t $ or the inertial -- $\omega _i $
frequencies. Findings are based on a well defined spectral peaks at
the sum -- $\omega _+ $ and dif -- $\omega_-$ frequencies of
inertial, semidiurnal and synoptic motions measured by current
meters and temperature records collected in the coastal Japan/East
Sea. We made rotary spectral analyses of these records and found
that the clockwise rotary spectrum for coastal currents measured at
35--m depth in the 1999 year has well-defined spectral peaks at the
frequencies $\omega_t \sim 1/12.4$(cph), $\Omega_{1,2} \sim
1/64,1/102$(cph), as well as sym and difference frequencies
$\omega_{\pm} =\omega_s \pm \Omega_{1,2} $ and their overtones
$\omega_n =n\omega_{\pm} (n=1,2,3)$. Analyses of coastal temperature
records collected in the 2004 year showed that the spectrum of
temperature variations has an analog form with spectral peaks at the
inertial frequency $\omega _{i} \sim 1/17.8$ (cph) and $\Omega
_{1,2} \sim 1/80,1/160$ (cph) synoptic frequencies, as well as sym
and difference frequencies $\omega_{\pm} =\omega _i \pm \Omega
_{1,2} $ and overtones $\omega_n =n\omega_{\pm} ( n=1,2,3)$.

With the theory of nonlinear interaction among week dispersion waves
[{\it Gurbatov et al.}, 1990] we consider the problem on the
parametric interaction of the nonlinear Kelvin wave at the frequency
$\omega$ and the low-frequency narrow-band nose with representative
frequency $\Omega\ll\omega $. We show that nonlinear interaction
between Kelvin wave and no stationary low-frequency component of the
background coastal current excited by the atmospheric forcing
leading to the intensification of tidal and inertial currents on the
sub-surfers and near-bottom layers of the coastal zone. Analogy
phenomena of the Kelvin wave intensification can occur in the
equatorial ocean.

\section{Observations}

\begin{figure}[t]
\noindent
\includegraphics[width=20pc]{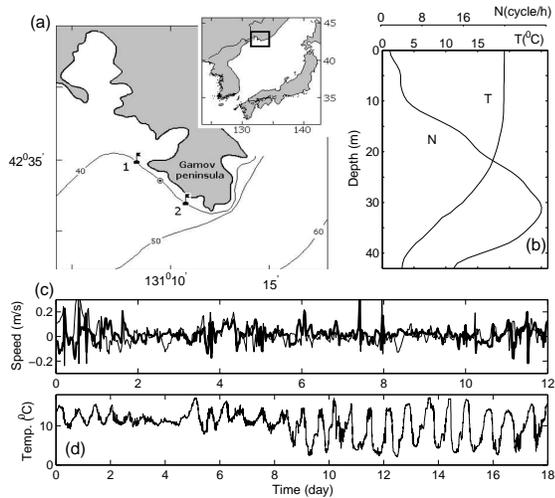}
\caption{(a) Study area on the Japan/East Sea shelf, September 1999,
2004. The mooring location in 2004 is shown in Arabic. The circle
indicates the mooring location in 1999 and the location of the
mooring vessel, from which hourly casts were conducted on September,
1999, 2004. (b) Daily mean vertical profiles of temperature $T$ and
buoyancy frequency $N$. (c) Records of the meridional (light curve)
and zonal (heavy curve) components of the current at the 35-m depth
in autumn 1999. (d) Record of the temperature variations at the
mooring buoy 1 in autumn 2004.}\label{fig1}
\end{figure}

In the first experiment, observations of internal waves were
performed at the coast of the Gamov Peninsula area of the Japan/East
Sea in 1999 year. The records of current meter collected from a
mooring deployed water depth 40 m, during 2 weeks in September
(Figure~\ref{fig1}a). To gauge temperature, speed and direction of
currents, the buoy was instrumented with the Russian-made POTOK
integral instrument at a depth of 35 meters. The POTOK had a
temperature measurement resolution of 0.05$^{\circ}$C. The sampling
rate was 15 minutes. The time series of the meridional (light curve)
and zonal (heavy curve) current components from this mooring are
shown on Figure~\ref{fig1}c. The Canadian Guideline CTD profiler was
used for the vertical profiling of water. The hydrostatic pressure,
temperature and electric conductivity of seawater were measured
during profiling. The profiling errors were no greater than
0.01$^{\circ}$C in temperature and 0.02 psu in salinity.
Figure~\ref{fig1}b shows the mean temperature and buoyancy frequency
profiles for the study site. Details of this experiment (results,
methods of their processing, etc.) were reported by {\it
Novotraysov, et. al} [2003].

In the second experiment, measurements of internal wave band
temperature fluctuations were performed during 18 days starting 3
September 2004. The time records of temperature were collected from
two moorings deployed along the coastline at a distance of 800m from
it, approximately at 40m depth and separated by a distance of 5.5 km
from each other. The first of them was deployed at 28m depth and the
second one, at 35m depth (below the surface). The moorings were
equipped with digital thermographs made by the Russian manufacturer.
The devices had a measuring precision of 0.05$^{\circ}$C for
temperature. The sampling rate was 1 minutes. The temperature and
salinity vertical profiles were performed on 20--21 September, from
a vessel anchored between the moorings with the Canadian Guideline
CTD profiler. In total, 25 hourly casts were made.

The obtained data were analyzed by unified standard spectral
techniques [{\it Emery and Thomson}, 1997]. The approach involved
(i) the elimination of low-frequency components with a Tukey's
cosine filter, (ii) the splitting of the resulting series into nine
37.2-hour segments (three semidiurnal tidal periods each), (iii) the
calculation and averaging of spectral densities by segments, and
(iv) the smoothing of the averaged spectral densities by a
five-point Tukey's filter. The number of degrees of freedom in the
processing amounted to approximately 10 for the first experiment and
20 for the second, providing a reasonable reliability of the results
of spectral analysis.

\section{The model}

A spectrum model of nonlinear interactions among internal Kelvin
waves is developed. It is assumed that $H/\lambda \ll 1$, and
$A/H\ll 1$, where $H$ is the water depth, $\lambda $ is a
characteristic wave-length, $A$ is a representative wave amplitude.
The basic component of this model is the simple wave equation. For
the first most powerful mode of the small-amplitude Kelvin waves the
equation is written as

\begin{equation}
\label{eq1} {\partial u}/{\partial x}-\alpha u{\partial u}/{\partial
\tau} =0,
\end{equation}
where $u(t,x)$ is the alongshore current, $x$ is a horizontal
coordinate, $\tau =t-x/c$, $t$ is time. The parameters $\alpha $ and
$c$ are the coefficient of nonlinearity and the phase speed of long
internal waves, respectively. Parameter of nonlinearity is
determined by the background density and is related in the
Boussinesq approximation as:

\begin{equation}
\label{eq2} \alpha =\left( {3H\int\limits_{-H}^0 {\phi _z^3 dz} }
\right)\times \left( {2c^2\int\limits_{-H}^0 {\phi _z^2 dz} }
\right)^{-1},
\end{equation}
where $z$ is a vertical coordinate, positive upward. The phase speed
$c$ and the amplitude function of the wave mode $\phi (z)$ are
determined from the solution of the eigenvalue problem

\begin{equation}
\label{eq3} d^2\phi {\kern 1pt}/dz^2+c^{-2}N^2(z){\kern 1pt}\,\phi
=0,
\end{equation}
with boundary condition

\begin{equation}
\label{eq4} \phi (-H)=\phi (0)=0,
\end{equation}
and with the normalization

\begin{equation}
\label{eq5} \phi _{\max } =1.
\end{equation}
Equation (\ref{eq1}) is a basic model for study of Kelvin waves
interaction. Let us consider interaction nonlinear Kelvin wave with
frequency $\omega_{0}$ (tidal $\omega_{t} $ or inertial $\omega_{i}
)$ and narrow-band synoptic noise (later SN) with frequency
$\Omega\ll\omega_{0}$ using the spectrum model of nonlinear internal
waves described in terms of the equation (\ref{eq1}). Let the
alongshore current $u(x,t)$ at the boundary of the coastal area
$x=0$ be the superposition of internal wave with frequency $\omega
_{0} $, amplitude $A_{0} $ and the noise $\vartheta(t)$ with typical
frequency $\gamma <<\omega_{0} $

\begin{equation}
\label{eq6} u(t,x=0)=u_0 (t)=A_{{\kern 1pt}0} \cos {\kern
1pt}\,(\omega _0 {\kern 1pt}t+\varphi )+\vartheta (t),
\end{equation}
where $\varphi$ -- is a random phase with uniform distribution in
the interval $[-\pi ,+\pi ]$.

We confine our analysis to the wave evolution stage, which is
characterized by condition $x<x_\ast $, where $x_\ast=\,(\alpha A_0
\omega_{0} )^{-1}$. In this stage the front Kelvin wave appears and
it is not accompanied by generation of internal solitons. We
introduce parameter $d=x/x_\ast $, which determine the similarity
between a Kelvin shock-wave and the wave and consider the case
$d<1$. For this case the spectral density of the wave field $u$ is

\begin{eqnarray}
\label{eq7} a(\omega;x)=-\frac{J_0(\omega d/\omega_0)}{2\pi i\omega
\alpha x} \int\limits_{-\infty }^{+\infty }\{\exp [-i\omega \alpha
x\vartheta (t)]-1\} e^{i\omega t}dt-\nonumber\\
-\sum\limits_{n=-\infty \atop n\ne 0}^\infty\frac{i^{n}J_{n}(\omega
d/\omega_0)}{2\pi i\omega \alpha x}\int\limits_{-\infty }^{+\infty
}{\{\exp [in\varphi-i\omega \alpha x\vartheta (t)]-1\}} \times \nonumber\\
\times \exp [i(\omega-n\omega_0)t]dt\,\,\,\,
\end{eqnarray}

From equation (7) we can affirm that the first term is the spectral
density of the synoptic oscillations and the second term is the
spectral density of the Kelvin waves distorted by the synoptic
oscillations.

We consider the spectral composition of the Kelvin wave near of the
harmonics with the number $n$ of the carrier frequency $\omega _0 $.
Given that $\gamma \ll \omega _{0}$ we make replacement $\omega $ at
$n\omega _0 $ in the exponential rate of the equation (7). Then we
obtain the equation for the harmonic with number $n$

\begin{equation}
\label{eq8} u_n \left( {t,x} \right)=A_n (x)\cos [n\omega_{0}
t+n(\varphi(t)-\omega_{0} \beta x\vartheta (t))]
\end{equation}
and her amplitude depending on the parameter $d=A_0 \alpha \omega_0
x$ is

\begin{equation}
\label{eq9} A_n(x)=2J_n(nd)A_0/nd.
\end{equation}
From the equation (\ref{eq9}), we can affirm that the amplitude of
the harmonic is decreased with increasing of harmonic number $\sim
1/n$.

\begin{figure*}[t]
\noindent
\includegraphics[width=39pc]{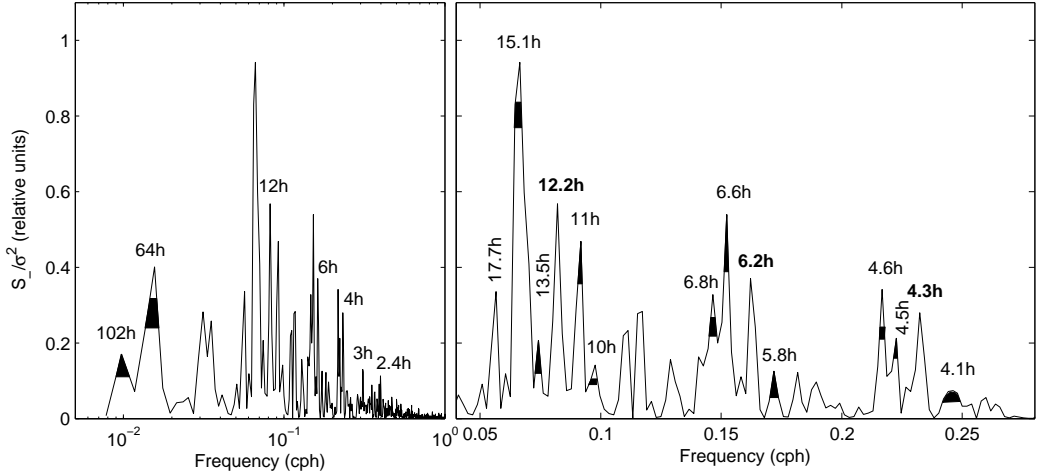}
\caption{(a) Normalized clockwise rotary spectra of currents versus
for current measured at 35--m depth in autumn 2004 near the Gamov
peninsula. The numerical digits over peaks are their periods (h).
(b) The increased fragment of the spectrum in the surrounding of the
semidiurnal frequency. Dark numerals {\bf 12.2; 6.2; 4.3} are
carrier periods (h).}\label{fig2}
\end{figure*}

Perform the more detailed spectral analysis of the nonlinear Kelvin
waves, when the synoptic noise consists of regular and stochastic
components: $\vartheta (t)=a\cos (\Omega t)+\vartheta _{ns} (t)$.
Let the amplitude of the phase modulation related with regular
component $M=\alpha n\Omega x$ is small, that is $M\ll 1$. Let the
Kelvin waves have narrow-band spectrum, and the synoptic noise width
$\gamma \ll \omega_{0} $, besides the noise, amplitude and phase of
Kelvin waves have Gaussian distribution. In this case the
correlation function of the harmonic with number $n$ taking into
account the equation (\ref{eq8}) is

\begin{eqnarray}
\label{eq10} B_n (\tau ,x)= \{\frac{A_n^2}{2}\{\cos n\omega_{0} \tau
+\frac{(a\alpha n\omega_{0} x)^2}{2}[\cos (n \omega _{0}
+\Omega\tau)+\nonumber\\
+\cos (n\omega_{0} -\Omega )\tau ] \} \}\exp [-n^2D_\psi \tau/2],
\,\,\,\,\,\,\,\,\,\,\,\,\,\,\,\,\,\,\,\,\,\,\,\,
\end{eqnarray}
where $D_\psi (\tau )=D_\varphi (\tau )+(\alpha \omega_{0}
x)^2(\sigma _{ns}^2 -B_{ns} (\tau ))$. Here $D_\phi $ is the
structure function of the Kelvin wave phase, $B_{ns} (\tau
)=<\vartheta _{ns} (t+\tau )\vartheta _{ns} (t)>$ is the correlation
function of the synoptic noise. We perform the Fourier transform of
the correlation function $B_n \left( {\tau ,x} \right)$ and obtain
the formula for the spectrum of the harmonic

\begin{eqnarray}
\label{eq11} S_n \left( {\omega ;x} \right)=\frac{A_n ^2}{2}\tilde
{S}_n \left( {\omega -n{\kern 1pt}\omega _0 }\right)+
\frac{(A_n a\alpha n\omega _{{\kern 1pt}0} x)^2}{4} \times \nonumber\\
\,\,\,\,\,\,\,\,\,\,\,\,\,\,\,\,\times \{\tilde {S}_n \left( {\omega
-(n{\kern 1pt}\omega _0 +\Omega } \right))+\tilde {S}_n (\omega
-(n{\kern 1pt}\omega _{{\kern 1pt}0} -\Omega ))\},
\end{eqnarray}
where

\begin{equation}
\label{eq12} \tilde {S}_n \left( {\omega \,;x}
\right)=\frac{1}{2\pi}\int\limits_{-\infty }^{+\infty } {\exp
[-n^2D_\psi (\tau )/2]\cos } {\kern 1pt}(\omega \tau )d\tau.
\end{equation}

From equation (\ref{eq11}), we can affirm that spectrum of the wave
have harmonics with frequencies $n\omega _{0} $ as well as harmonics
with sum and dif frequencies $\omega_{\pm}=n\omega_{0} \pm \Omega $
and amplitudes increasing with growth of distance traversed by the
wave. This means that at the coastal zone between Kelvin waves and
synoptic noise takes place nonlinear interaction, which is
accompanied by intensification Kelvin waves with side frequencies
$\omega_{\pm}=n\omega_{0} \pm \Omega $.

\section{Discussion and Conclusion}
\begin{table*}[t]
\caption{The periods of the peaks of the temperature variations
spectrum (digit numerals) and the calculated periods (light
numerals)}\label{tab1}
\begin{tabular}{|c|c|c|c|c|c|c|c|c|c|}
\hline
& & & & & & & & & \\
$1/\Omega(h)$ & $1/\omega_{+}(h)$ & $1/\omega_{s}(h)$ &
$1/\omega_-(h)$ & $1/\omega_{+}(h)$ & $1/\omega_{i}(h)$ &
$1/\omega_{-}(h)$ & $1/\omega_{+}(h)$ & $2/\omega_{i}(h)$ &
$1/\omega_{-}(h)$ \\
& & & & & & & & & \\
\hline \hline
& & & & & & & & & \\
84& 10.8& 12.4& 14.5& 14.7& 17.8& 22.6& 8.0& 8.9&10.0
\\
& & & & & & & & & \\
\hline
& & & & & & & & & \\
\textbf{84}&\textbf{11.0}&\textbf{12.4}&\textbf{14.5}&
\textbf{15.0}&\textbf{17.5}&\textbf{22.5}&\textbf{8.0}&
\textbf{8.7}& \textbf{10.0} \\
& & & & & & & & & \\
\hline
& & & & & & & & & \\
168& 11.5& 12.4& 13.4& 16.1& 17.8& 19.9& 8.5& 8.9& 9.4
\\
& & & & & & & & & \\
\hline
& & & & & & & & & \\
\textbf{168}& \textbf{11.5}& \textbf{12.4}& \textbf{13.0}&
\textbf{16.2}& \textbf{17.5}& -- & \textbf{8.5}& \textbf{8.7}&
\textbf{9.5}\\
& & & & & & & & & \\
\hline
\end{tabular}
\end{table*}
Let us turn to the records of the current velocity.
Figure~\ref{fig1}c presents records smoothed by Tukey window with
the width 1 h of the meridional (light curve) and zonal (heavy
curve) components of the current in the coastal zone of Japan/East
Sea in autumn 1999 year. It is of interest to note a low-frequency
component of the variation in the record of the current velocity.
\begin{figure}[t]
\noindent
\includegraphics[width=20pc]{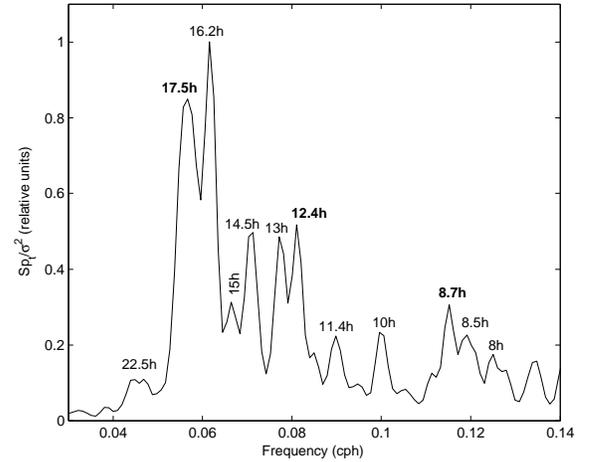}
\caption{Normalized spectrum versus   for record of the temperature
variations at the mooring buoy 1 in autumn 2004 near the Gamov
peninsula. The numerical digits over peaks are their periods
(h).}\label{fig3}
\end{figure}
Let us analyze circular rotary current component with the rotary
spectral analysis [{\it Emery and Thomson}, 1997]. The spectrum of
the clockwise rotary (CWR) component of the velocity is

\[
S_-(\omega )=0,125\cdot (S_{uu} +S_{vv} -2Q_{uv} ),
\]
where $S_{uu} $ and $S_{vv} $ are the one-sided autospectra of the
$u$ and $v$ Cartesian components of the velocity and $Q_{uv} $ is
the quadrature spectrum between the two components. These spectra
were determined by unified statistical spectral techniques using the
algorithms presented in [{\it Emery and Thomson}, 1997]. The spectra
have N degrees of freedom, where N $\ge $ 10. Figure~\ref{fig2}a
shows the CWR spectrum of currents versus $\log {\kern 1pt}\,\omega
$ normalized by maximal value. It is of interest to note the groups
of the well defined significant spectral peaks in the surrounding of
the semidiurnal frequency and her $1,2,3$ and $4$ harmonics. Our
estimations show that the magnitudes of these spectral peaks
interrelated with number harmonics by the approximate relation

\[
A_{12} :A_6 :A_4 :A_3 :A_{2,4} \approx 1:2:3:4:5
\]
(with error $<$ 20{\%}).

There are two peaks with light and heavy tops at the representative
frequencies $\Omega _1 \sim 1/64$ and $\Omega _2 \sim 1/102$ (cph)
in the low-frequency band of the spectra.

Figure~\ref{fig2}b presents the increased fragment of the spectrum
in the surrounding of the semidiurnal frequency. It is of interest
to note a significant well defined spectral peak at the $\omega_{s}
=1/12.2$ (cph) frequency surrounded by side peaks with dark end
light tops. Not hard to make sure that the frequencies of these
picks are determined by formula $\omega _\pm =\omega _{{\kern 1pt}s}
\pm \Omega _{1,2} $ with error equals of experimental error
uncertainty. In particularly the strong spectral peak at the
difference frequency $\omega _- =1/12.2-1/64$ (cph) suggests that
the motion in study region is dominated by not only the tide
movements, but and the inertial movements. The pick at the frequency
$\omega _{i} =1/17.7$ (cph) is formed by the inertial movements.
Thus the spectrum has fine structure in the band 1/14 - 1/16.5
(cph), which don't segregate with help our dates and methods of
analyses.

Let us consider the structure of the spectrum in surroundings of the
harmonics of semidiurnal frequency $\omega _{s}$
(Figure~\ref{fig2}b). As in the past picks at the carrier
frequencies: $\omega _{s1},\omega_{s2} $ are surrounded by side
picks at the frequencies, whish are determined by formula $\omega
_\pm =\omega _{s1,2} \pm \Omega _{1,2}$.

Thus spectral analysis of the clockwise rotary velocity shows that
her spectrum is determined by oscillations with the tidal frequency
$\omega _{s} \approx 1/12.2$ (cph) and the frequencies $\Omega
_{1,2} \approx 1/72,106$ (cph) from synoptic band and also that
spectrum has the fine structure in the neighborhood of the frequency
$\omega _s $ with side peaks at the frequencies, which are
determined by formula $\omega _\pm =\omega _s \pm \Omega _{1,2} $
The similar structure has spectrum in the neighborhood of the first
and the second harmonics of the semidiurnal frequency $\omega _{s}
$.

Then turn to the records of the temperature of the coastal water of
Japan/East sea collected in autumn 2004 year. Figure 3 shows the
averaged spectrum of the temperature records collected from two
moorings and normalized by the maximum value of the spectrum.
Spectra are calculated by unified standard spectral techniques [{\it
Emery and Thomson}, 1997]. The number of freedom degrees amounted to
approximately 20.

Of particular interest are the groups of the significant spectral
peaks at the inertial frequency and her first harmonic and also the
peaks in the surrounding of the semidiurnal frequency $\omega _s $.

Arabic numerals over spectral peaks equal the periods (hour) of
these peaks. These periods (h) are placed in the Table~\ref{tab1}.
Digit numerals equal the periods of the spectral peaks and light
numerals equal the periods calculated by the formula $1/\omega _\pm
=1/\omega _{s,i} \pm 1/\Omega _{1,2} $.

As indicated by Table~\ref{tab1} the periods of the spectral peaks
and the periods calculated by the formula have approximately equal
values. Thus the spectral analysis of current and temperature meter
records of the Japan/East Sea coastal water shows that the spectrum
is determined by oscillations with frequencies of near inertial
$\omega _i \approx 1/17.8$ (cph) and tidal $\omega _{s} \approx
1/12.2$ (cph) frequencies as well as the frequencies $\Omega _{1,2}
\approx 1/72,106$ (cph) from synoptic band. The analysis shows that
the spectrum has the fine structure in the neighborhood of the
frequencies $\omega _{i,s} $ with side picks at the frequencies,
which are determined by formula $\omega_\pm =\omega _{i,s} \pm
\Omega _{1,2}$. The similar structure has spectrum in the
neighborhood of the first harmonics of the inertial frequency
$\omega _{i}$.

As a conclusion, we showed that among nonlinear internal Kelvin wave
with frequency $\omega$ and the low-frequency narrow-band nose with
representative frequency $\Omega\ll\omega $ occur the parametric
interaction, i.e. among the low-frequency band and high-frequency
band of internal Kelvin waves occur the energy exchange. Hence, as a
result of nonlinear interaction the energy of internal Kelvin waves
in the coastal and tropical oceans can grow at the expense of the
low-frequency narrow-band nose energy.

\begin{acknowledgments}

This work was supported by the Presidential Grant No.
MK--1364.2008.5 by the Prezidium of the Russian Academy of Sciences
(Program ``Mathematical methods in nonlinear dynamics.'' Project No.
06-I-13-048).

\end{acknowledgments}

\end{article}

\end{document}